\documentclass[a4paper,11pt]{article}
\usepackage{pos}
\usepackage{booktabs}

\title{Master-field simulations of QCD}

\notes{~\\[2em]\flushright CERN-TH-2021-198}

\emailAdd{fritzscp@tcd.ie}
\author*[a]{Patrick~Fritzsch}
\author[b]{John~Bulava}
\author[c]{Marco~Cè}
\author[d,c]{Anthony~Francis}
\author[c,d]{Martin~Lüscher}
\author[e,c]{Antonio~Rago}

\affiliation[a]{School of Mathematics, Trinity College Dublin, Dublin 2, Ireland}
\affiliation[b]{John von Neumann-Institut für Computing (NIC), DESY, Platanenallee 6, 15738 Zeuthen, Germany}
\affiliation[c]{Department of Theoretical Physics, CERN, 1211 Geneva 23, Switzerland}
\affiliation[d]{Albert Einstein Center for Fundamental Physics, Institute for Theoretical Physics, Universität Bern, Sidlerstrasse 5, 3012 Bern, Switzerland}
\affiliation[e]{Centre for Mathematical Sciences, Plymouth University, Plymouth, PL4 8AA, United Kingdom}

\hypersetup{pdfauthor={P.~Fritzsch, J.~Bulava, M.~Cè, A.~Francis, M.~Lüscher, A.~Rago},pdftitle={Master-field simulations of QCD}}

\abstract{%
We report on the first master-field simulations of QCD with 2$+$1 dynamical
quark flavours using non-perturbatively improved stabilised Wilson fermions.
Our simulations are performed at a lattice spacing of 0.094 fm with 96 and 192
points in each direction.  On both lattices, the pion and kaon masses are equal
to 270 and 450 MeV, respectively, and $m_{\pi} L$ thus reaches an unprecedented
value of $25$ on the larger lattice.  This setup matches a single point on a
chiral trajectory with fixed trace of the quark mass matrix and allows for
comparisons to standard large-scale simulations.  We present our algorithmic
setup and performance measures, and report about our experience in thermalising
large master-field lattices with fermions.
}

\FullConference{%
 The 38th International Symposium on Lattice Field Theory, LATTICE2021
  26th-30th July, 2021
  Zoom/Gather@Massachusetts Institute of Technology
}

\newcommand{\brad}[1]{\langle\kern-3pt\langle{#1}\rangle\kern-3pt\rangle}

\begin{document}
\maketitle

\section{Introduction}

The idea that a very large lattice gauge field, the \emph{master field}, can
replace a conventional Markov chain Monte-Carlo ensemble of a discretised field
theory with a mass gap, has been brought forward at the Granada Lattice
conference~\cite{Luscher:2017cjh}.  If the physical size of the lattice is
large enough, master-field simulations provide a solution to the long-standing
topology-freezing problem which severely affects standard simulations when the
lattice spacing is decreased.  Beside this technical aspect, master fields
allow to explore new kinematical regimes and to address new physics questions
which could not be addressed in the standard framework. On the other hand, not
every observable that can be computed efficiently with well-established methods
may be suitable in master-field calculations.  For a statistical data analysis
on a master field, the usual ensemble average gets replaced by a translation
average of localised observables computed in physically distant regions of
space-time.  Beside the numerical evidence provided in
ref.~\cite{Luscher:2017cjh}, this method has been applied successfully in a
pure $SU(3)$ gauge theory calculation of the topological susceptibility above
the critical temperature~\cite{Giusti:2018cmp}. Including (especially light)
quarks to the simulation is challenging as the generation of these fields using
standard techniques leads to various algorithmic instabilities and precision
issues. Ways to overcome these problems have been described in
ref.~\cite{Francis:2019muy} for the O($a$)-improved Wilson formulation of
lattice QCD. As most of these stability measures are generically applicable,
they are also successfully used in typical large-scale simulations nowadays,
cf.~\cite{Lattice2021:266,Lattice2021:277}.

In these proceedings, we report on the first master-field simulation(s) of QCD
by reviewing our computational setup in the next section and reporting about
our experiences with simulations of such very large lattices. While we will
focus on the technical aspects here, first physics results are presented in a
separate contribution to these proceedings~\cite{Lattice2021:228}.

\section{Master-field simulations}

The computational setup for master-field simulations derives from our initial
study of stabilising measures for QCD simulations with Wilson
fermions~\cite{Francis:2019muy}.  We use the tree-level Symanzik-improved gauge
action~\cite{Weisz:1982zw} and 2$+$1 dynamical quark flavours employing the
new, exponentiated $O(a)$-improved Wilson fermion action with
non-perturbatively tuned Sheikholeslami--Wohlert coefficient $c_{\rm sw}$.  The
A-lattices simulated in ref.~\cite{Francis:2019muy} constitute an approximate
chiral trajectory at a rather coarse lattice spacing of $a=0.094\,$fm
($\beta=6/g_0^2=3.8$). It enables us to approximately fix $m_{\pi}=270\,$MeV
and $m_{\rm K}=450\,$MeV in advance. We aim at simulating master fields of size
$96^4$ and $192^4$ with periodic gauge and anti-periodic fermion fields.  In
this way we will be able to study the statistical accuracy of physical
observables and possible systematic effects through various sub-volume
averages. These lattices correspond to $m_\pi L=12.5$ and $25$, or lattice
extents of $9\,$fm and $18\,$fm, respectively.
All simulations are performed with the publicly available \texttt{openQCD}
code~\cite{Luscher:openQCD} which supports master-field simulations from
version 2.0 on. It also supports parallel IO and quadruple precision arithmetic
in global sums, both particularly relevant for very large lattices.

\subsection{Algorithm details}

Instead of the usual Hybrid Monte-Carlo (HMC) algorithm~\cite{Duane:1987de}, we
use the Stochastic Molecular Dynamics (SMD)
algorithm~\cite{Horowitz:1985kd,Jansen:1995gz} to update the gauge fields
$U(x,\mu)$, pseudo-fermion fields $\phi(x)$ and momentum fields $\pi(x,\mu)$.
Each update cycle starts with a refreshment of $\pi$ and $\phi$ fields (keeping
$U$ constant) using a linear combination of their current state with random
fields drawn from a normal distribution. Subsequently, the molecular dynamics
(MD) equations are integrated over a time-distance $\epsilon$ using a
reversible symplectic integration rule, followed by the usual accept-reject
step to guarantee that the update cycle
$(U,\pi,\phi)\to(U^{\prime},\pi^{\prime},\phi^{\prime})$ preserves the correct
distribution.  If a new field configuration is rejected, we restart with
$(U,-\pi,\phi^{\prime})$, i.e., the gauge field is set to $U$ and the initial
momentum field to $-\pi$ while keeping $\phi^{\prime}$ fixed. For sufficiently
small $\epsilon$, it can be shown that this SMD algorithm is
ergodic~\cite{Luscher:SMDnote}. In practise, we aim at an average acceptance
rate $\langle P_{\rm acc}\rangle = \langle\min\{1,e^{-\Delta H}\}\rangle\ge
98\%$ which, at the lattice spacings considered in~\cite{Francis:2019muy},
typically leads to values of $\epsilon< 1/3$.  While such high acceptance rates
are usually avoided when the HMC algorithm is in use, it is the relevant range
when using the SMD algorithm. Beside $\epsilon>0$, the SMD introduces a
friction parameter $\gamma>0$, which determines how quickly the memory of
previous field configurations is lost.  Throughout our simulations, we fix it
to $\gamma=0.3$ which amounts to a choice that was found to optimize the
observed autocorrelation times of physical quantities in $SU(3)$ gauge
theory~\cite{Luscher:2011kk}.%
\footnote{In contrast to the discussion in~\cite{Luscher:2017cjh}, it was
        emphasized in~\cite{Francis:2019muy} that the effort of simulating
        HMC and SMD is comparable such that the algorithmic exactness, using the
        accept-reject step (more frequently for the SMD), should not be given up.
}%

Apart from the algorithm described above, we chose established techniques in
our simulation: even-odd preconditioning, SAP (Schwarz alternating procedure)
preconditioned deflated solvers, twisted-mass factorisations for the
light-quark determinant, rational approximations for the strange-quark
determinant and a hierarchical 4th-order integrator with two integration levels
for the MD equations.  To exclude statistically relevant effects of numerical
inaccuracies while simulating very large lattices, it is advantageous to
slightly adapt the solver parameters for each of the following three
categories. A uniform-norm stopping criterion is used in the generation of
pseudo-fermion fields and the corresponding force calculations, while the
standard two-norm is used in the action computations. We furthermore set the
bound on the residues to $10^{-12}$ for the solvers applied to the action and
pseudo-fermions, while those for the forces derived from the pseudo-fermions of
the light-quark doublet are set to $10^{-11}$ and those of the heavier strange
quark to $10^{-10}$.

\subsection{Thermalisation strategy}

\begin{figure}[t!]
        \centering
       \includegraphics[width=0.8\textwidth,page=1]{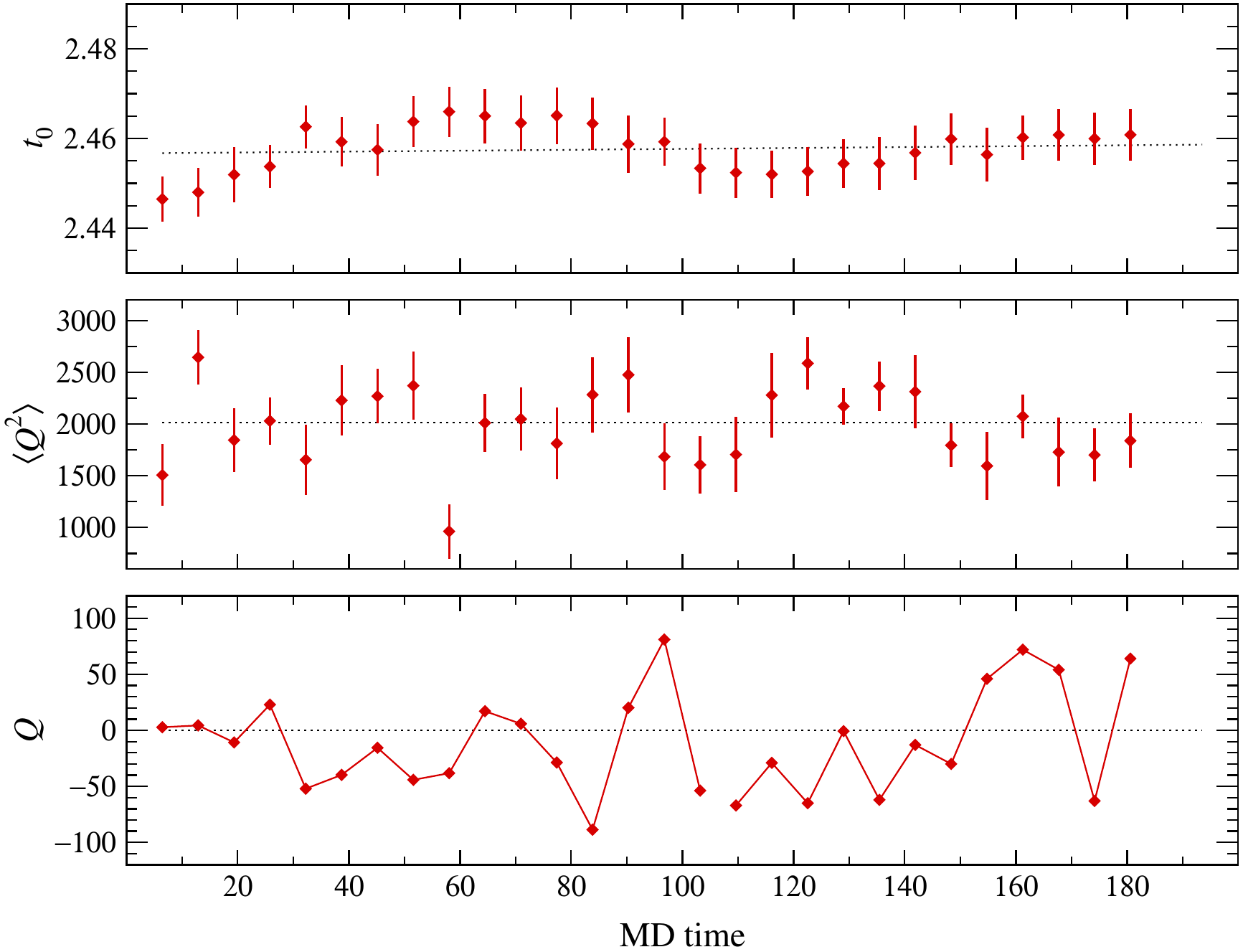}\\[0.8em]
       \;\includegraphics[width=0.7\textwidth,page=1]{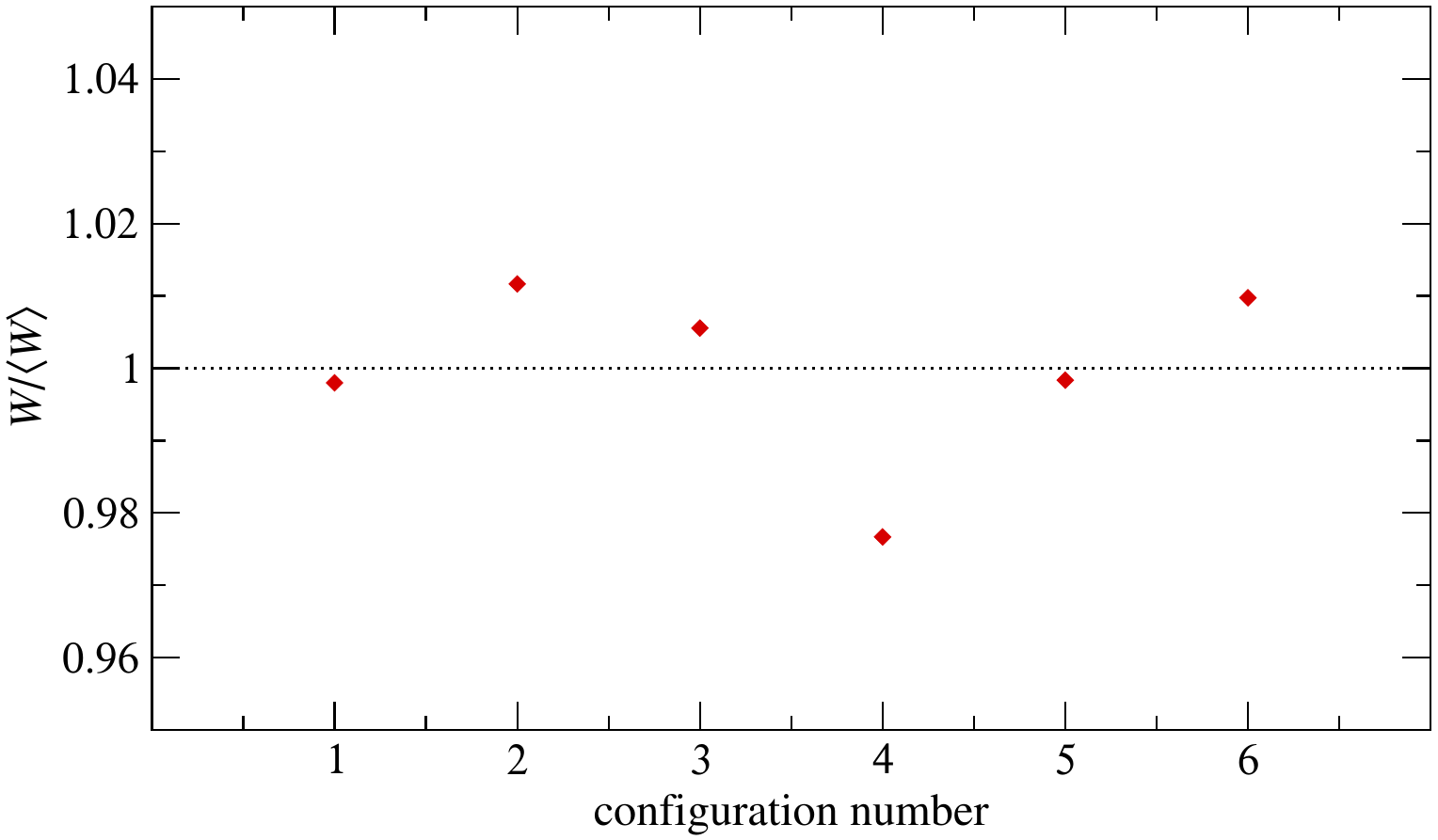}\\
       \caption{Monitoring observables during final thermalisation of the $96^4$ lattice.}
        \label{fig:Obs}
\end{figure}

The cost of simulating master-field-sized lattices essentially agrees with the
computing time required to reach thermal equilibrium. This effort mainly scales
with some power of the volume and the exponential autocorrelation time of the
underlying algorithm. To date, both remain an obstacle for speeding up
large-scale simulations of QCD. Hence, we follow common thermalisation
strategies, i.e., we start from smaller lattices and periodically extent one
direction at a time to allow the gauge field to relax to the new setup and to
adapt algorithmic parameters, if needed. Our first goal was to reach thermal
equilibrium on the $96^4$ lattice before moving on to $192^4$.  We started from
an $A_2$ configuration with small total topological charge $Q$ (a $96\cdot
32^3$ lattice with $m_{\pi}=294\,$MeV, cf.~\cite{Francis:2019muy}), and changed
the light and strange quark hopping parameters to our target masses as well as
the twisted-mass reweighting parameter $\mu_{0}=0.002 \to 0.0$. For master
fields, the latter is required as standard reweighting techniques are not
compatible with the way expectation values are being calculated, cf.
refs.~\cite{Luscher:2017cjh,Lattice2021:228}.  But what about the (Zolotarev
optimal) rational approximation of the strange-quark determinant which
generically requires reweighting? In this case, the reweighting corrects for
the numerical approximation error that can be reduced with increasing
simulation cost. In practise, one has to ensure that the approximation error is
negligible for physical observables $O$. 
The absolute deviation between the reweighted and non-reweighted observable
fulfills the following generic and strict bound~\cite{Luscher:openQCD-doc-rhmc} 
\begin{align}
        \Big| \frac{\langle OW \rangle}{\langle W\rangle}-\langle O \rangle \Big| &\le \sigma_{O}\frac{\sigma_{W}}{\langle W\rangle} \;.
\end{align}
Accordingly, any observable $O$ is guaranteed to be unbiased w.r.t. reweighting
if, for instance,  ${\sigma_{W}}/{\langle W\rangle} \le 0.1$ is chosen as upper
bound on the relative uncertainty of the reweighting factors. Here $\sigma_W$
is the standard deviation of the strange-quark reweighting factors $W$, and
$\sigma_{O}$ that of the observable. For this reason, it is advisable to also
compute the strange-quark reweighting factors on the smaller lattices, beside
typical observables used to monitor the ongoing thermalisation phase.
In fig.~\ref{fig:Obs} we show a subset of our monitoring observables during the
final thermalisation of the $96^4$ lattice: the flow time $t_0$ (computed with
Wilson flow from the clover energy density), the total topological charge
$Q=\sum_{x} q(x)$, and the topological susceptibility $\langle
Q^{2}\rangle/V\simeq\sum_{|x|\le 20}\brad{q(x)q(0)} $ as a function of the MD
time. Note that $\brad{\cdot}$ denotes the aforementioned sub-volume average,
cf.~\cite{Luscher:2017cjh}.  The latter two have been measured directly at flow
time $t=2.46$.  Uncertainties have been estimated through master-field
translation averages.  These quantities have large integrated autocorrelation
times and are therefore expected to be among the observables that thermalise
most slowly. In table~\ref{tab:Cost96} we present our cost figure for reaching
the first $96^4$ master field. The chosen rethermalisation times are
significantly larger than the expected autocorrelation time of 30 MDU and any
residual thermalisation effects are therefore expected to be well below the
statistical errors.
\def\hph{\hphantom{1}}
\begin{table}
        \small
        \centering
\begin{tabular}{lrrrrr}\toprule
             lattice        & \#cores       & $t_\mathrm{SMD}[\mathrm{sec}]$ 
                                                     & $t_\mathrm{MDU}[\mathrm{sec}]$ 
                                                                  & MDU   & Mcore-h     \\\midrule
     $96^{\hph}\cdot  32^3$ & $ 16\cdot 48$ &  246   &      794   & 155   &       0.03  \\
     $96^2   \cdot  32^2$   & $ 48\cdot 48$ &  277   &     1108   & 125   &       0.09  \\
     $96^3   \cdot  32$     & $ 64\cdot 48$ &  672   &     2800   & 176   &       0.42  \\
     $96^4$                 & $128\cdot 48$ &  1080  &     5020   & 206   &       1.77  \\\midrule
     total:                 &               &        &            & 662   &       2.31  \\
  \bottomrule
\end{tabular}
\caption{Thermalisation cost till first $96^4$ master-field lattice. We provide
        the measured average time per SMD cycle ($t_\mathrm{SMD}$) and the associated
        time to complete one molecular dynamic unit ($t_\mathrm{MDU}$), followed
        by the simulation length (in MDU) for each lattice size and the corresponding   
        simulation cost in million core-hours.
        }
        \label{tab:Cost96}
\end{table}
In all thermalisation steps we additionally checked for spikes in $\Delta H$
and that $\langle e^{-\Delta H}\rangle=1$ holds within errors. No problems have
been observed.  The residual systematic effects of the rational approximation
are controlled by adjusting the approximation such that $\sigma_W$ is 1-2\% of
its average value. On the $96^4$ lattice, and with the spectral range of
[0.015,8.0], this level is reached with 11 poles.

\subsection{Eigenvalues of the deflated Dirac operator}

\begin{figure}[h!]
        \vspace*{-1em}
        \centering
        \includegraphics[width=0.82\textwidth]{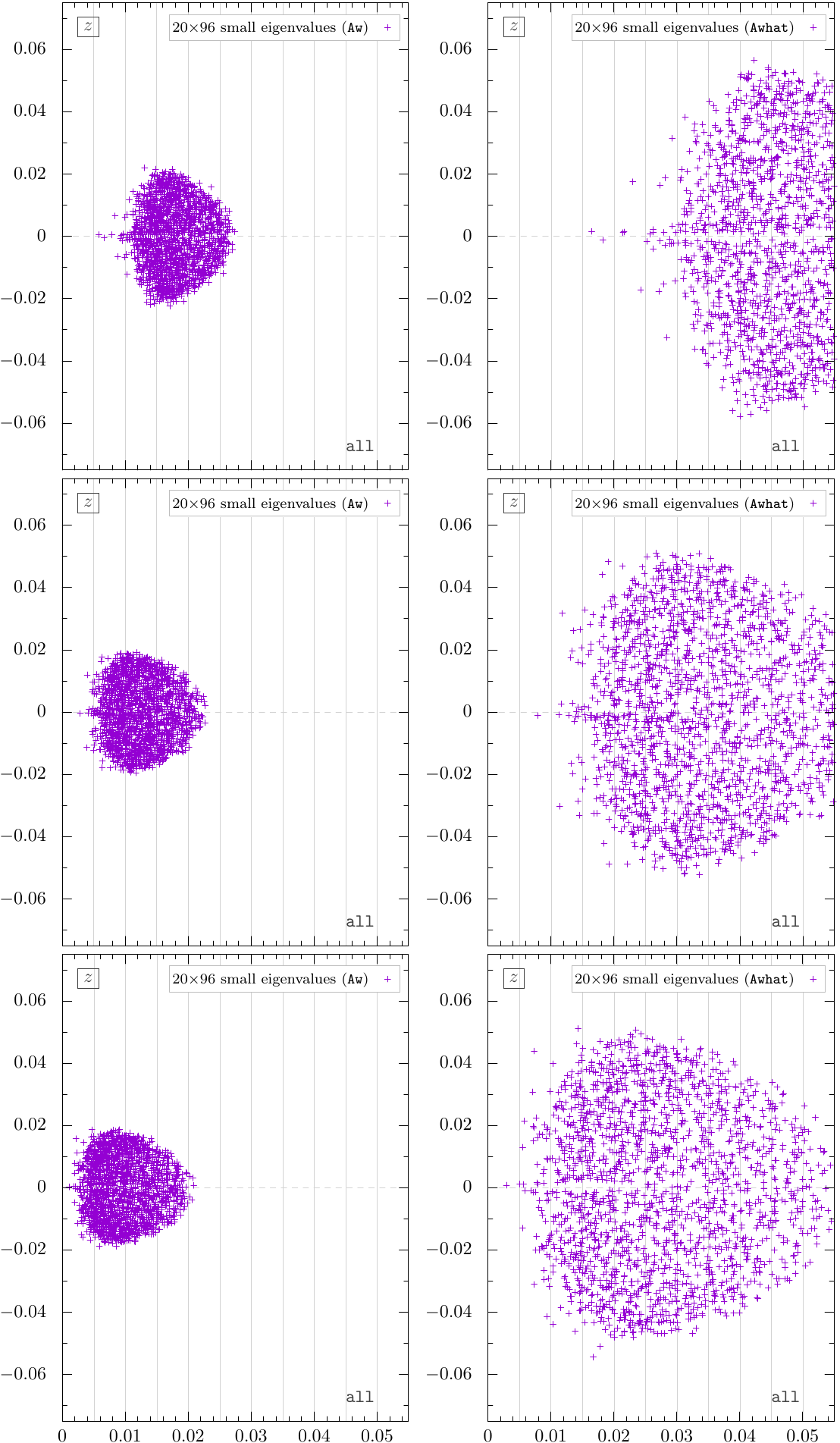}
        \caption{Lowest (complex) eigenvalue spectrum of the little Dirac operator
                $A_\mathrm{w}$ (\emph{left}) and $\hat{A}_\mathrm{w}$
                (\emph{right}) for light pseudoscalar meson masses of $408$,
                $293$ and $215$ MeV (top to bottom) on a $96\times 32^{3}$
                lattice ($a=0.094$\,fm). The data points comprise the lowest 96
                eigenvalues computed on each of 20 configurations. 
        }
        \label{fig:AwAwhat}
\end{figure}

In the very same way, we continued increasing the volume towards our $192^4$
master field. With increasing $V$ we began to observe rare spikes in $\Delta H$
as well as failures to update, generate or regenerate the deflation subspace.
This can happen for no particular reason, because the little system can become
nearly singular even if the Dirac operator is well conditioned. As a protective
measure, the deflation (DFL) subspace was more regularly regenerated and the
number of Krylov vectors for the solver of the little (deflated) system was
significantly increased, both to no avail. The number of problematic events
increased with the volume (and smaller pion masses), impacting the performance
and continuity of the simulation. Its source is mostly unknown and multiple,
interfering effects could play a role. To investigate the issue we computed the
eigenvalues of the little Dirac operator, $A_{\rm w}=P_0DP_0$, using the freely
available software package
\texttt{SLEPc}~\cite{Hernandez:2005:SSF,petsc-efficient,petsc-user-ref,petsc-web-page}
for this non-hermitean operator. $P_0$ is an orthogonal projector to the DFL
subspace, spanned by a set of global low-modes
$\mathscr{L}_D=\{\psi_1,\dots,\psi_{N_s}\}$ of the Dirac operator. For more
details we refer to refs.~\cite{Luscher:2007se,Frommer:2013fsa}.  The results
are presented in fig.~\ref{fig:AwAwhat} for both the smallest (magnitude)
eigenvalues of $A_{\rm w}$ and its even-odd preconditioned version
$\hat{A}_{\rm w}$ on our set of $A$-lattices. They obviously confirm the
general expectation that the smallest eigenvalue decreases with the quark mass
for both operators. Computing the 96 lowest eigenvalues on a total of 20
configurations, we can clearly identify a few eigenvalues close to the real
axis which are about a factor 2 smaller than the majority that forms an
approximate boundary wall. Such fluctuations can result from the roughness of
the gauge field  or from unaccounted numerical instabilities of unknown source
(after all we are simulating at coarse lattice spacing, $a=0.094\,$fm).
Although we were always able to restart the simulation from reseeding the
random number generator, it requires additional measures to continuously
simulate large master-field lattices.

\subsection{Multilevel deflated solver}

To solve the problem, a multilevel deflated solver is being introduced as a
natural extension of the current single-level DFL solver of
\texttt{openQCD-2.0}, and effectively preconditions the standard little Dirac
operator. For that purpose, a stack of DFL subspaces is  introduced for
block-grid levels $0\le k\le k_{\rm max}$, such that the little Dirac operator
at block-level $k$ fulfills $A_{k}=P_{k}A_{k-1}P_{k}=P_kDP_k$. This means that
$A_{\rm w}=A_0$ if $k_{\rm max}=0$ and that all block-level operators can be
derived from the same set of global low modes $\mathscr{L}_D$. Especially
larger lattices, able to host two or more deflation levels, can profit from the
additional hierarchy through reduced computational costs. In contrast to the
standard single-level DFL solver, the new implementation uses double precision
arithmetic also in the projection/lifting operations to further gain stability
at the expense of an additional overhead. We continued thermalising the $192^4$
master field with this new solver and indeed observed a significant reduction
of failing solves. The question remains whether further tuning of the solver
parameters can resolve the remaining instabilities. In fig.~\ref{tab:Cost192}
we also provide a cost figure for thermalising the $192^4$ master field. Due to
various parameter and algorithmic changes we only give average time estimates,
corresponding to the true cost in core-hours. 

\begin{table}[t!]
        \small
        \centering
\begin{tabular}{lrrrrr}\toprule
             lattice        & \#cores       & $\bar t_\mathrm{SMD}[\mathrm{sec}]$ 
                                                     & $\bar t_\mathrm{MDU}[\mathrm{h}]$ 
                                                             & MDU    & Mcore-h    \\\midrule
     $192^{\hph}\cdot 96^3$ & $128\cdot 48$ & 2740   &  794  &  95    &   2.32     \\
     $192^2     \cdot 96^2$ & $256\cdot 48$ & 3080   & 4.73  &  45    &   2.54     \\
     $192^3     \cdot 96$   & $512\cdot 48$ & 3190   & 5.34  &  35    &   4.49     \\
     $192^4$                & $768\cdot 48$ & 4789   & 9.71  & 102    &  35.12     \\\midrule
     total:                 &               &        &       & 277    &  44.47     \\
  \bottomrule
\end{tabular}
\caption{Thermalisation cost till first $192^4$ master-field lattice.}
        \label{tab:Cost192}
\end{table}

\section{Summary}

In this work, we have described our experience simulating master-field lattices
of 2$+$1 flavour QCD with non-perturbatively improved exponentiated-clover
Wilson fermions.  The previously proposed stabilising measures clearly improve
our ability to simulate coarse lattice spacings and large lattices at the same
time. Very large physical-volume simulations, like $(18\,\textrm{fm})^4$, have
become possible using today's high-performance computers but remain challenging
for Wilson fermions with anti-periodic boundary conditions. To further mitigate
failures of the underlying solvers, we introduced the multilevel deflated
solver, allowing us to produce a single $192^4$ master-field lattice.  It
appears that the required gain in stability comes at additional costs,
confirming the no-free-lunch theorem once more. In the future we plan to
perform new master-field simulations at a finer lattice spacing and a smaller
quark mass. Going forward, we plan to further study the problems encountered at
coarse lattice spacing and look forward to first physics applications on the
reported master fields.

\acknowledgments

We acknowledge PRACE for awarding us access to SuperMUC-NG at GCS@LRZ, Germany.
The eigenvalue measurements were performed on a dedicated HPC cluster at CERN.
This project has received funding from the European Union’s Horizon 2020
research and innovation programme under the Marie Skłodowska-Curie grant
agreement No. 843134. 

\providecommand{\href}[2]{#2}\begingroup\raggedright\endgroup

\end{document}